\documentclass[%
reprint,
superscriptaddress,
amsmath,amssymb,
aps,
]{revtex4-2}

\usepackage{xcolor}
\usepackage{graphicx}
\usepackage{dcolumn}
\usepackage{bm}

\begin{document}

\preprint{APS/123-QED}

\title{Six-wave mixing of optical and microwave fields using Rydberg excitations in thermal atomic vapor}

\author{Tanim Firdoshi}
\email{tanim.firdoshi@niser.ac.in}
\author{Sujit Garain}
\author{Suman Mondal}
\author{Ashok K. Mohapatra}
\email{a.mohapatra@niser.ac.in}
\affiliation{National Institute of Science Education and Research Bhubaneswar, Jatni 752050, India}
\affiliation{Homi Bhabha National Institute, Training School Complex, Anushaktinagar, Mumbai 400094, India}

\date{\today}

\begin{abstract}
Rydberg EIT-based microwave sensing has limited microwave-to-optical conversion bandwidth due to fundamental limitation in the optical pumping rate to its dark state. We demonstrate a parametric six-wave mixing of optical probe and coupling fields driving the atoms to a Rydberg state via two-photon excitation and two microwave fields with frequency offset of $\delta$ driving the Rydberg-Rydberg transition in thermal atomic vapor. Microwave-to-optical conversion bandwidth of $17$ MHz is achieved in the present experiment which is limited by the available coupling power. Further theoretical investigation of the system presents higher modulation bandwidth with larger coupling Rabi frequency.
\end{abstract}

\maketitle

Wireless network architecture has been developing swiftly with the aid of atomic systems. The role of atoms in the quest for development of quantum technology \cite{kimb08}, sensing electric \cite{sedl12,sedl13,fan15,simo16} and magnetic fields \cite{savu05,kosc10} has engrossed researchers to explore certain atomic properties. High sensitivity, higher dynamic range, self-calibrating ability, and the advantage of being traceable to international standards are some major aspects that lead to the dominance of atomic sensors over traditional electric field sensors \cite{holl14,holl17,meye21}. Wide operational frequency range from MHz to THz and long lifetime of highly excited Rydberg atoms \cite{gall94} can enable the coupling of microwave and THz fields to optical fields \cite{wade17} leading to applications such as imaging of millimeter waves \cite{gord14} and sub-wavelength imaging \cite{fan14,hollo14}, quantum states transfer from microwave to optical domain and storage of quantum information \cite{hafe12}. Most of the experiments involve the use of cavities \cite{rued16,andrews14,strek09} to enhance the coupling of microwaves to optical fields. Rydberg atoms due to the large electric dipole transitions can enable free space coherent conversion of microwave fields to optical photons which has been observed via a six-wave mixing process in a cold atomic system \cite{vogt19,han18} using electromagnetically induced transparency (EIT) \cite{moha07}. There has also been a study on using EIT for digital communication with Rydberg atoms\cite{meye18}. The EIT probing scheme restricts the data transfer rate to the order of a few Mbit/s~\cite{meye18,deb18}. This rate is limited fundamentally owing to the optical pumping rate to the EIT dark state. The need for wireless network with faster data transfer rate leads to the exploitation of Rydberg atoms with widely available atomic transitions.
\begin{figure}
	\begin{center}
		\includegraphics[width=8.5cm]{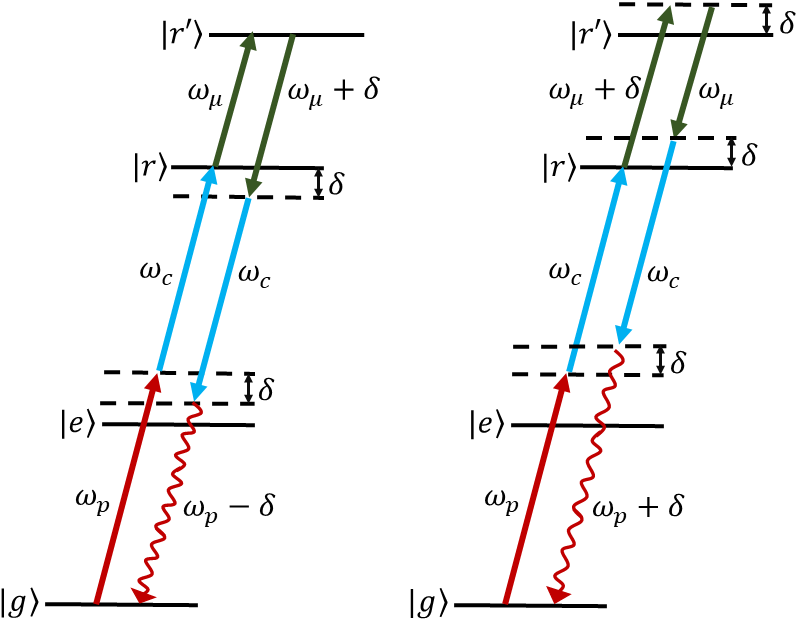}
		\caption{\label{fig1} Schematic of the energy level diagram of four-level atomic system. The probe laser field with frequency $\omega_p$ and detuning $\Delta_p$ couples the transition $|g\rangle \longrightarrow|e\rangle$. The coupling laser with frequency $\omega_c$ and detuning $\Delta_c$ couples the transition $|e\rangle \longrightarrow|r\rangle$. The transition between two Rydberg states $|r\rangle \longrightarrow |r'\rangle$ is coupled by two microwave fields with frequencies $\omega_{\mu}$ and  $\omega_{\mu}+\delta$.}
	\end{center}
\end{figure}

In this letter, we demonstrate the parametric generation of new optical frequencies due to six-wave mixing of optical and microwave fields using coherent Rydberg excitation in thermal rubidium vapor. We also present a theoretical model of four-level system based on perturbative expansion of the density matrix to support the experimental observations. Due to the parametric nature of the process, microwave-to-optical frequency conversion rate is not fundamentally limited unlike the optical pumping rate to the dark state in the case of Rydberg EIT. We demonstrate the microwave-to-optical conversion bandwidth to be $17$ MHz which is limited by the available coupling laser power in our experiment. We further present a theoretical investigation of the same system to increase the bandwidth using suitable laser parameters. This work pave the way to use Rydberg atom based  sensor for faster data transfer in microwave communication technology. Thermal vapor system is also attractive due to its simplicity in technological implementation of this quantum device.

\begin{figure}
	\begin{center}
		\includegraphics[width=8.5cm]{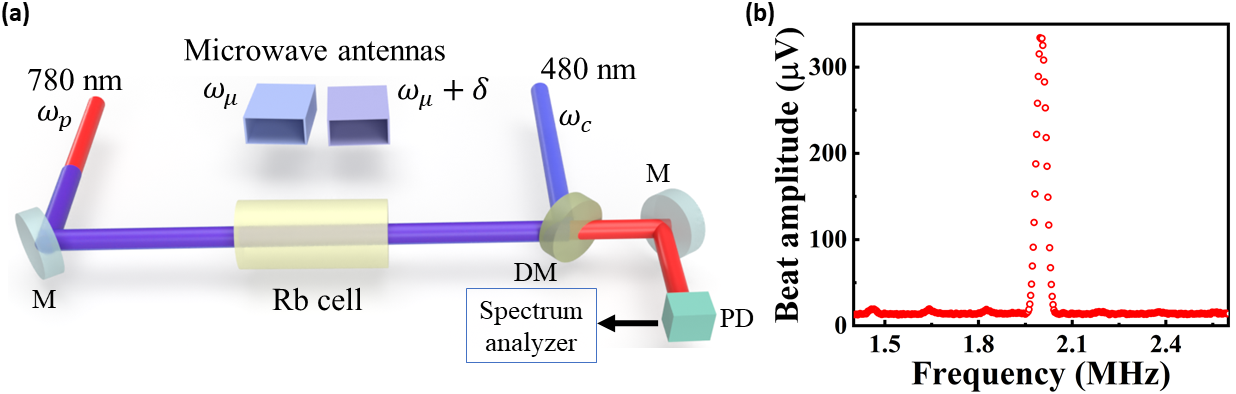}
		\caption{\label{fig2} (a) Schematic of the experimental set-up for the observation of six-wave mixing process. M: Mirror, DM: Dichroic mirror, PD: Photo-detector. (b) Spectrum analyzer signal showing the 2 MHz interference beat signal of the probe and the generated fields.}
	\end{center}
\end{figure}
Six-wave mixing process is understood by considering a four-level atomic system as shown in Fig. \ref{fig1}. The photon with frequency $\omega_p+\delta$ $\left(\omega_p-\delta\right)$ is parametrically generated by the $\chi^{(5)}$ process due to absorption of the optical photons with frequnecies $\omega_p$ and $\omega_c$ and a microwave photon with frequency $\omega_{\mu}+\delta$ $\left(\omega_{\mu}\right)$ followed by emission  of a microwave photon with frequency $\omega_{\mu}$ $\left(\omega_{\mu}+\delta\right)$ and an optical photon with frequency $\omega_c$. Assuming that the wave vectors of the microwave fields to be small, the phase matching of the process allows the generated field to copropagate with the $\omega_p$-photon. In a suitable rotating frame, the Hamiltonian of the system is $\hat{H}=-\hbar[\Delta_{p}|e\rangle\langle e|+(\Delta_{p}+\Delta_{c})|r\rangle\langle r|+(\Delta_{p}+\Delta_{c}+\Delta_{\mu})|r'\rangle\langle r'|]-\frac{\hbar}{2}[\Omega_{p}|g\rangle\langle e|+\Omega_{c}|e\rangle\langle r|+(\Omega_{\mu_{1}}+e^{-i\delta t}\Omega_{\mu_{2}})|r\rangle \langle r'|+H.c.]$ where $\Delta_{p}$ ($\Omega_{p}$), $\Delta_{c}$ ($\Omega_{c}$) and $\Delta_{\mu}$ ($\Omega_{\mu_{1}}$, $\Omega_{\mu_{2}}$) are the detunings (Rabi frequencies) of the probe, coupling and microwave fields respectively. The Rabi frequencies are defined as $\Omega_{p}=2\mu_{ge}A_{p}/\hbar$ , $\Omega_{c}=2\mu_{er}A_{c}/\hbar$, $\Omega_{\mu_{1}}=2\mu_{rr'} A_{\mu_{1}}/\hbar$ and $\Omega_{\mu_{2}}=2\mu_{rr'}A_{\mu_{2}}/\hbar$ where $\mu_{ge}$, $\mu_{er}$ and $\mu_{rr'}$ are the dipole moments corresponding to $|g\rangle\rightarrow|e\rangle$, $|e\rangle\rightarrow|r\rangle$ and $|r\rangle\rightarrow|r'\rangle$ transitions respectively. $A_{p}$, $A_{c}$, $A_{\mu_{1}}$ and $A_{\mu_{2}}$ are the amplitudes of the probe, coupling, strong microwave and the weak microwave fields respectively. Without loss of generality, all the fields are considered to be real. The optical Bloch equations for the system is given by $\dot{\rho}=\frac{i}{\hbar}[\rho,\hat{H}]+\mathcal{L_D}(\rho)$ where $\rho$ is the density matrix of the system and $\mathcal{L_D}(\rho)$ is the Lindblad operator which includes the decay and decoherence processes in the system \cite{lind76}. $\Gamma_{eg}$, $\Gamma_{re}$ and $\Gamma_{r'r}$ denote the population decay rates for the decay channels $|e\rangle\rightarrow|g\rangle$, $|r\rangle\rightarrow|e\rangle$ and $|r'\rangle\rightarrow|r\rangle$ respectively. The transit time decays of the thermal atoms in $|r\rangle$ and $|r'\rangle$ states through the cross section of the beam are given by $\Gamma_{rg}$ and $\Gamma_{r'g}$ and are taken to be $4.2$ MHz considering crossection of the applied laser beams. The decay rates used in the model for the calculations are $\Gamma_{eg}=6$ MHz, $\Gamma_{re}=\Gamma_{r'r}=0.01$ MHz. 

$\Omega_{\mu_{2}}$ is considered to be weak as compared to $\Omega_{\mu_{1}}$ and density matrix equations are solved in the steady state using a perturbative expansion as reported in \cite{bhow16}. The density matrix is written as $\rho=\rho^{(0)}+\rho^{(1)}e^{-i\delta t}+\rho^{(-1)}e^{i\delta t}$+(higher order terms), where $\rho^{(0)}$ is the zeroth order density matrix element oscillating with frequency $\omega_{p}$, $\rho^{(1)}$ and $\rho^{(-1)}$ are the first order elements that oscillate with frequency $(\omega_{p}+\delta)$ and $(\omega_{p}-\delta)$. The zeroth order equations are solved numerically which are used to solve the first order equations in the steady state and hence, calculate the first order coherence between ground and excited state $\rho^{(\pm1)}_{eg}$. The first order equations can be solved by neglecting the second order and higher order terms as the model is based on the consideration that one of the microwave field is weak. The polarization of the probe field $P(\omega_{p})\propto\rho^{(0)}$ and the generated field is $P(\omega_{p}\pm\delta)\propto\rho^{(\pm1)}$. The Doppler averaged susceptibility for the thermal atoms is $\chi^{(5)}_{eff}(\omega_{p}\pm\delta)=\frac{N\mid\mu_{ge}\mid}{\epsilon_{0}A_{p}A_{c}^{2}A_{\mu_{1}}A_{\mu_{2}}\sqrt{2\pi}v_p}\int_{-\infty}^{+\infty}\rho^{(\pm1)}_{eg}e^{-v^2/2v_p^2}dv$ where $v_p$ is the most probable speed of the atoms and $N$ is the vapor density.

The probe and the generated fields are detuned from the resonance with $\Delta_{p}=1.2$ GHz which leads to negligible absorption of the probe and generated fields. Using the slowly varying amplitude approximation and assuming the absorption of the generated fields to be small, the wave propagation equation for the generated field is written as $\frac{dA_{\pm1}}{dz}=\kappa_{\pm1}$ where $\kappa_{\pm1}=-i\frac{5}{2}\frac{k_{\pm1}}{n_{\pm1}^{2}}\chi_{eff}^{(5)}(\omega_{p}\pm\delta)A_{p}A_{c}^{2}A_{\mu_{1}}A_{\mu_{2}}$ is the non-linear coupling coefficient corresponding to the generated frequencies and it contains the susceptibility due to the six-wave mixing process. The solution to the propagation equation is given by $A_{\pm1}=\kappa_{\pm1}l$ where $l$ is the length of the vapor cell.

The schematic of the experimental setup is presented in Fig. \ref{fig2}(a). An external cavity diode laser operating at 780 nm is the probe field with frequency $\omega_{p}$ and a frequency doubling cavity laser operating at 480 nm is the coupling field with frequency $\omega_{c}$. The two beams counter propagate each other and are focused at the center of the rubidium vapor cell. $1/e$-radius of the probe (coupling) laser beam is $32$ $\mu$m ($63$ $\mu$m). The peak Rabi frequency at the waist of the probe (coupling) beam used in the experiment is 283 (5) MHz. The probe laser frequency is locked at 1.2 GHz blue detuned from the transition $^{85}$Rb $5S_{\frac{1}{2}}$ $F=3$ $\rightarrow$ $5P_{\frac{3}{2}}$ $F=4$. The vapor cell was heated to about $100$ $^0$C to achieve the density of $5\times10^{12} \text{cm}^{-3}$. The coupling laser is scanned around $\Delta_{p}$ to satisfy the two-photon resonance to the Rydberg state $|52D_{\frac{5}{2}}\rangle$. Two synchronized microwave signal generators are used to apply microwave fields through two horn antennas such that they propagate in the horizontal plane through the center of the vapor cell. The strong microwave field $\omega_{\mu}$ is applied at resonance frequency of 15.0895 GHz which couples the $|52D_{\frac{5}{2}}\rangle\longrightarrow|53P_{\frac{3}{2}}\rangle$ transition. A weak microwave field with a frequency offset $\delta$ couples the same transition. The Rabi frequency of the strong microwave field is $80$ MHz and the weak microwave field strength is varied between 1 MHz to 40 MHz. The six-wave mixing process generates the optical fields at frequencies $\omega_p\pm\delta$. The newly generated fields interfere with the probe field to give the beat signal at frequency $\delta$ which is detected using a fast photo-detector and spectrum analyzer as shown in Fig. \ref{fig2}(b). At resonance, the beat signal is represented as $A_{0}\cos(\delta t+\phi)$ where $A_{0}$ is found out to be $A_{0}=2A_{p}A_{+1}\frac{\sin(\phi_{+}+\phi_{-})}{\sin(\phi+\phi_{-})}=2A_{p}A_{-1}\frac{\sin(\phi_{+}+\phi_{-})}{\sin(\phi_{+}-\phi)}$ and $\phi=\frac{A_{+1}\phi_{+1}-A_{-1}\phi_{-1}}{A_{+1}+A_{-1}}$. $\phi_{+}$ and $\phi_{-}$ are the phases acquired by the generated fields due to the $\chi^{(5)}$ process which is very small. 
\begin{figure}
	\centering
	\includegraphics[width=8.5cm]{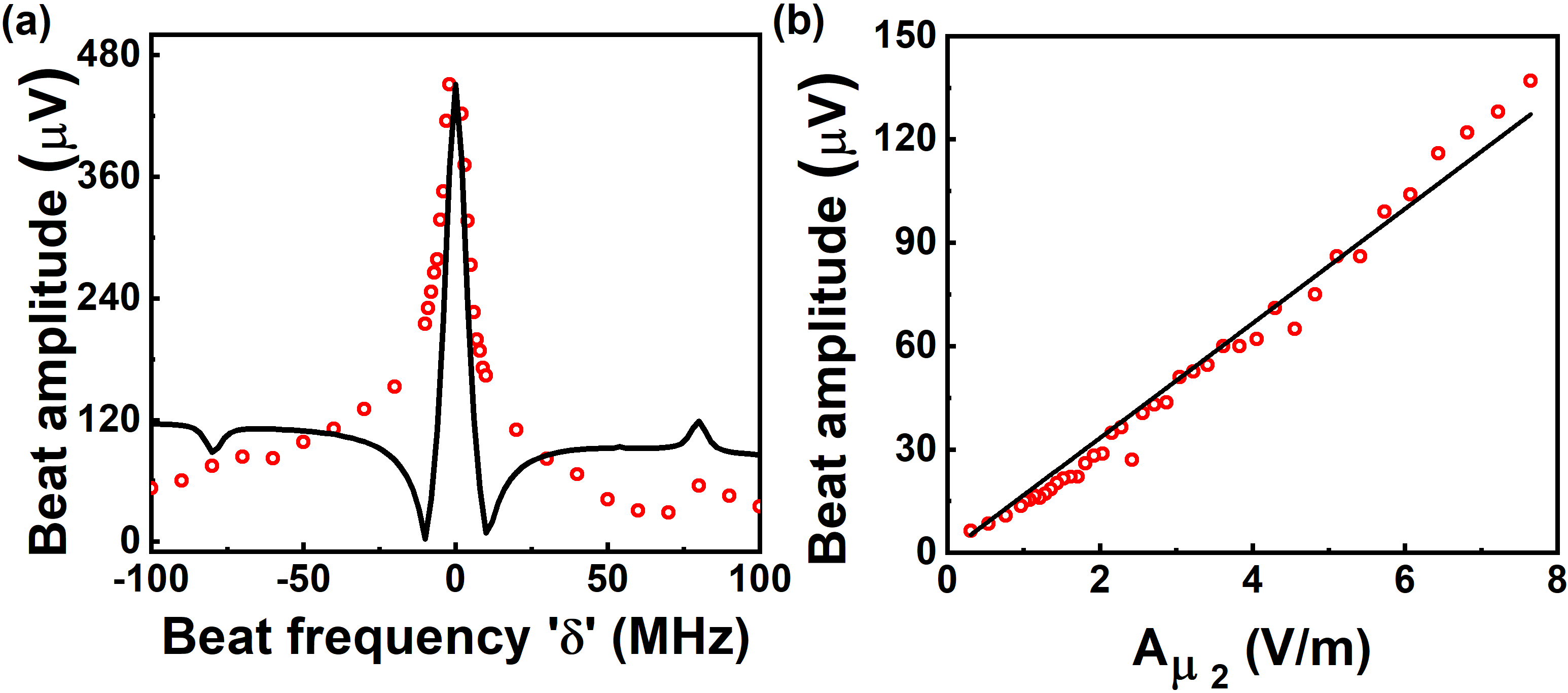}
	\caption{\label{fig3} (a) Beat signal amplitude with varying $\delta$ where the field $\omega_{\mu}$ is on resonance and the field $\omega_{\mu}\pm\delta$ is scanned around resonance. Red open circle represents the experimental data and the black solid line represents the fitting with the theoretical model. (b) Beat signal amplitude with varying weak microwave field strength at $\delta=10$ MHz with $\Delta_{\mu}=0$. Red open circles are the experimental data and the black solid line is the straight-line fit to the data given by $y=a*A_{\mu_2}$ where $a=16.64 \pm 0.22$ is the gain in the system.}
\end{figure}

We study the generated field by observing the beat signal at frequency $\delta$. With the strong field $\omega_{\mu}$ on resonance, the frequency of weak microwave field $\omega_{\mu}+\delta$ is changed by varying $\delta$ to see the strength of the generated field as shown in Fig. \ref{fig3}(a). We observe that the generation of the new optical field is maximum around the resonance and the strength reduces away from the resonance with an FWHM of $\sim$ 17 MHz. Similar behavior is observed from the theoretical model. A large peak is observed around the resonance and two small peaks are observed at $\delta=\pm80$ MHz as expected due to the dressing of the transition by the strong microwave field with a Rabi frequency of 80 MHz. It is seen from the experimental data that the left side of the spectrum is broader than the right side of the spectrum as it includes the contribution from the $|52D_{\frac{3}{2}}\rangle$ which is nearly 80 MHz away from $|52D_{\frac{5}{2}}\rangle$ state \cite{moha07}. The contribution from the other nearby Rydberg states is not taken into account in the theoretical model. The bandwidth of the generation spectrum and the strength of the generated field are limited by the available coupling Rabi frequency. The bandwidth is larger as compared to the spectrum as observed for EIT systems \cite{han18}. We also study the dependence of the generated field on the applied weak microwave field intensity with all the other field parameters kept constant as shown in Fig. \ref{fig3}(b). The beat strength increases linearly with the E-field strength of the weak microwave field as expected from the wave propagation equations.

\begin{figure*}[!ht]
\centering
		\includegraphics[width=17cm]{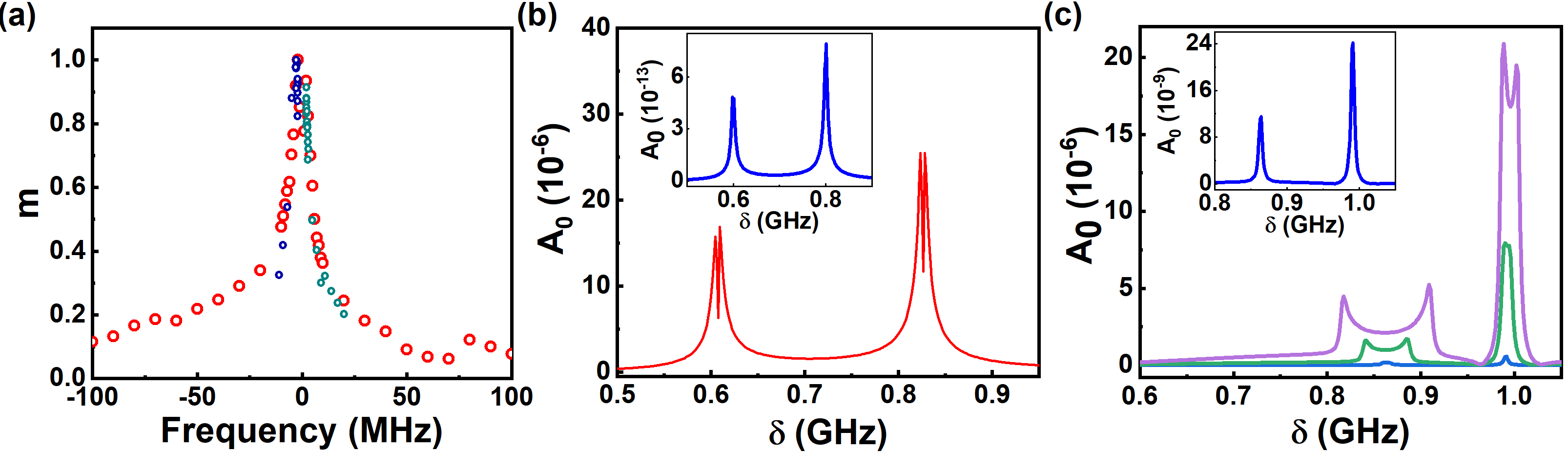}
		\caption{\label{fig4} (a) Comparison of the modulation spectrum with the spectrum of observed beat signal. Red open circles represent the normalized beat spectrum given by Fig. \ref{fig3}(a). Cyan open circles and blue open circles represent modulation index for the upper and lower side band of the modulated signal respectively at $\delta=2$ MHz. (b) Theoretically generated spectrum of $A_{0}$ for atoms at rest as function of $\delta$ with $\Omega_{p}=300$ MHz, $\Omega_{c}=100$ MHz, $\Omega_{\mu_{1}}=80$ MHz and $\Omega_{\mu_{2}}=1$ MHz. Inset shows the spectrum for $\Omega_{p}=10$ MHz, $\Omega_{c}=2$ MHz, $\Omega_{\mu_{1}}=10$ MHz and $\Omega_{\mu_{2}}=1$ MHz. (c) Theoretically generated spectrum of $A_{0}$ for thermal atoms as function of $\delta$ with increasing $\Omega_{c}$ of 10 MHz (blue line), 50 MHz (green line) and 100 MHz (purple line). Inset shows the spectrum for $\Omega_{c}=2$ MHz. Other parameters for this figure are $\Omega_{p}=300$ MHz, $\Omega_{\mu_{1}}=80$ MHz and $\Omega_{\mu_{2}}=1$ MHz.}
\end{figure*}

Communication technology employs the technique of amplitude modulation (AM)/frequency modulation (FM) of baseband signals onto an electromagnetic carrier and the data transfer rate is estimated from the modulated signal bandwidth \cite{jiao19}. The data transfer process in the EIT system involves the transfer of the modulation of microwave field to optical field whereas our system investigates the electro-optic conversion of the microwave field to the optical field via a parametric six-wave mixing process which is expected to be faster. We introduce a modulation $m(t)=A_{0}cos(\nu t)$ into the weak microwave field, where $\nu$ is the modulation frequency and $A_{0}$ is the amplitude of the modulation wave. This results in amplitude modulation of the generated field due to the six-wave mixing phenomenon, i.e., the generated field acts as a carrier of the modulation. Modulation of the weak microwave field gives rise to the generation of side-bands with frequencies given by $(\omega_{p}\pm\delta)\pm\nu$. Experimentally, we observe the side-bands at $(\delta\pm\nu)$ for the carrier at $\delta$. The amplitude of the side-bands is derived to be $mA_{p}(A_{+1}+A_{-1})$, where $m$ is the modulation index with $m\le1$. The value of $m$ is determined from the ratio of the amplitude of side-bands and carrier beat amplitude as a function of $\nu$ which is depicted in Fig. \ref{fig4}(a). We compare the bandwidth of the modulation spectrum with the spectrum observed for the generated beat signal presented in Fig. \ref{fig3}(a). We observe that the bandwidth of both the spectrum is in good agreement with each other. In EIT systems, the EIT pumping rate limits the signal bandwidth to up to 4 MHz \cite{deb18,han18}. In our experiment, the signal bandwidth is limited to 17 MHz because of the limitation in the available coupling laser power rather than any fundamental limitation as for the case of the EIT system.

We study theoretically the effect of the coupling Rabi frequency on the generated beat spectrum. We observe the beat spectrum at off-resonance condition, i.e. $\Delta_{p}=1200$ MHz, $\Delta_{c}=-600$ MHz and $\Delta_{\mu}=200$ MHz. In order to understand the observation in the thermal system, we first studied the system for atoms at rest with similar parameters as the thermal system. If we have low Rabi frequencies of laser fields, then we observe two peaks at $\delta=600$ MHz and $\delta=800$ MHz as shown in the inset of Fig. \ref{fig4}(b). Higher Rabi frequencies result in the increase in the strength of the signal as well as shifting of peaks due to light shifts which is shown in Fig. \ref{fig4}(b). Also, there is splitting of the peaks due to the formation of dressed states. Similarly, we observe two peaks for the thermal vapor system which are shifted due to the wave-vector mismatch and light shift factors as shown in the inset of Fig. \ref{fig4}(c). We observe that the spectrum is broadened with the increase in the coupling Rabi frequency, as shown in Fig. \ref{fig4}(c), which in turn can result in a modulation spectrum of larger bandwidth. If we observe the left side peak centered at around 864 MHz, $A_{0}$ increases by 240 times with the increase in coupling Rabi frequency from 2 MHz to 100 MHz along with the increase in the spectrum bandwidth to 100 MHz. It is to be noted that $A_{0}$ is comparable to the beat amplitude in the present experiment. The power required to achieve an average coupling Rabi frequency of 100 MHz is nearly 27 Watt with $\frac{1}{e}$-radius of the coupling beam to be 50 $\mu$m. The above observation is a strong indication that such a system can be highly useful in achieving a higher data transfer rate for the purpose of communication technology.

In conclusion, we have investigated the six-wave mixing using Rydberg atoms in a thermal rubidium vapor. We have demonstrated the parametric generation of new optical frequency from efficient mixing of optical and microwave fields. The experimental results are well supported by a theoretical model of a four-level system using the density matrix equations. Our system fulfills the most important requirement for microwave communication technology, i.e. larger modulation bandwidth, resulting in a faster data transfer rate. 

\begin{acknowledgements}
The authors thank Dr. S. Bedanta, Dr. K. Senapati and Rohde and Schwarz, India for providing us with microwave signal generators for the experiment. The authors also thank Dr. G. S. Babu for technical help related to the microwave antennas. The authors gratefully acknowledge the financial support from the Department of Atomic Energy, Government
of India under the Project Identification No. XII-R$\&$D-5.02-0200 (National Institute of Science Education and Research Bhubaneswar).
\end{acknowledgements} 

{}

\end{document}